# Ligand-induced oligomerization of the human GPCR neurotensin receptor 1 monitored in living HEK293T cells

Anika Westphal[a], Hendrik Sielaff[b], Stefanie Reuter[a], Thomas Heitkamp[b], Ralf Mrowka[a], Michael Börsch[b,c,*]

[a]Experimental Nephrology Group, Jena University Hospital, Nonnenplan 2 - 4, 07743 Jena
[b]Single-Molecule Microscopy Group, Jena University Hospital, Nonnenplan 2 - 4, 07743 Jena,
[c]Center for Medical Optics and Photonics (CeMOP) Jena, Germany

**ABSTRACT**

The human neurotensin receptor 1 (NTSR1) is a G protein-coupled receptor that can be expressed in HEK293T cells by stable transfection. Its ligand is a 13-amino-acid peptide that binds with nanomolar affinity from the extracellular side to NTSR1. Ligand binding induces conformational changes that trigger the intracellular signaling processes. Recent single-molecule studies revealed a dynamic monomer – dimer equilibrium of the receptor *in vitro*. Here we report on the oligomerization state of the human NTSR1 in the plasma membrane of HEK293T cells *in vivo*. We fused different fluorescent marker proteins mRuby3 or mNeonGreen to the C-terminus of NTSR1 and mutated a tetracysteine motif into intracellular loop 3 (ICL3) for subsequent FlAsH labeling. Oligomerization of NTSR1 was monitored before and after stimulation of the receptor with its ligand by FLIM and homoFRET microscopy (i.e. Förster resonance energy transfer between identical fluorophores detected by fluorescence anisotropy), by colocalization microscopy and by time-lapse imaging using structured illumination microscopy (SIM).

## 1 INTRODUCTION

Binding of extracellular small molecules to membrane-bound G protein-coupled receptors (GPCR) triggers intracellular signaling processes. Because the large superfamily of GPCRs is involved in distinct physiological processes, for example neurotransmission, behavioral regulation, immune response, taste, smell and vision, they are important targets for more than 30% of all therapeutic drugs applied today[1].

The neurotensin receptor 1 (NTSR1) belongs to the β group of class A GPCRs and modulates dopaminergic systems, analgesia and inhibition of food intake in the brain and digestive processes in the gut[2, 3]. NTSR1 is activated by the 13-amino-acid peptide neurotensin which binds with nanomolar affinity[4]. Ligand binding likely induces conformational changes in NTSR1 that trigger $G_q$ protein signaling processes as well as interactions with kinases and arrestin molecules within minutes. Lipid composition also affects signaling of NTSR1[5].

NTSR1 is one of few GPCRs which can be expressed in bacterial membranes, purified and reconstituted in liposomes. The X-ray structure of NTSR1 with bound ligand has been solved by R. Grisshammer and coworkers in 2012[6] enabling site-directed mutations for specific fluorescence labeling. Recently transient dimerization of purified NTSR1 in a solid-supported artificial lipid bilayer system has been reported[7] by single-molecule FRET (smFRET) measurements between differently tagged receptors. We have published experimental evidence for dimers of purified NTSR1 after reconstitution into liposomes[8] as found previously for detergent-solubilized receptors[9]. We evaluated the monomeric, dimeric and oligomeric state of this NTSR1 from rat labeled with mNeonGreen at its C-terminus using confocal time-resolved single-particle homoFRET measurements, both in detergent solution as well as after reconstitution into liposomes. However, it remains unclear whether the observed partial dimerization or oligomerization of reconstituted NTSR1 in the absence of its ligand neurotensin is a purification or reconstitution artefact, or the receptor exists as a monomer or a dimer or even oligomer in its native lipid environment in mammalian cells[10].

..................................................................................................................................
* email: michael.boersch@med.uni-jena.de; http://www.single-molecule-microscopy.uniklinikum-jena.de

Therefore we constructed a human NTSR1 with either mRuby3 or mNeonGreen fused to the C-terminus as a fluorescent marker for cellular imaging in living HEK293T cells. In addition we introduced a tetracysteine motif in NTSR1 for specific labeling with the bisarsenical fluorescein derivative FlAsH[11]. Here we investigated the ligand-induced redistribution and aggregation of NTSR1 in the membrane by widefield microscopy, structured illumination microscopy (SIM), fluorescence lifetime microscopy (FLIM) and anisotropy imaging. We detected homoFRET indicating coexistence of both the monomeric and the dimeric or oligomeric state of NTSR1 in the membrane of living HEK293T cells.

## 2 EXPERIMENTAL PROCEDURES

**2.1 Plasmid construction**

NTSR1 comprises 7 transmembrane helices, an extracellular N-terminus and an intracellular C-terminus[12, 13]. Two sequences in the third intracellular loop ICL3 were identified for possible replacement with the minimal tetracysteine motif CCPGCC[14] for binding of the bisarsenical fluorescent dye FlAsH[11, 15, 16]. Of the two sequences $_{272}$CTVGGE$_{277}$ and $_{290}$IEPGRV$_{295}$ the replacement of the latter with the tetracysteine motive was tested for successful FlAsH labeling here. At the C-terminus the sequence for the fluorescent protein mRuby3[17] was inserted. The donor plasmid with the combined genes was introduced into human HEK293T FlpIn cells to generate a stable cell line for homogenous gene expression and protein production. Similarly we generated a plasmid for the human NTSR1 with mNeonGreen[18] at the C-terminus.

The DNA sequence of the codon-optimized mutant NTSR1 gene, together with a minimal cytomegalovirus (CMG) promoter (CMG-NTRS1), was synthesized (Eurofins Genomics, Ebersberg, Germany) and ligated into the multiple cloning site (MCS) of the vector pEX-A128 (Eurofins Genomics, Ebersberg, Germany) using appropriate restriction enzymes.

In preparation for cloning a specific MCS for the integration of the NTSR1 construct and mRuby3 was integrated into the modified safe harbor vector AAV-CAGGS-EGFP (Addgene # 22212), which was a gift of Rudolf Jaenisch[19]. This vector contains two homology arms for the integration of the promoter of interest into a safe harbor region located in the AAVS1 gene.

Primers with appropriate overhangs (forward: gctattaattaaggtcaccaccatcatcaccatg, reverse: tgcacttaagtGTCTCGAGCTATTACTTGTACAGCTC) were designed and used for amplification of mRuby3 from a plasmid containing the synthesized sequence for mRuby3 (Eurofins Genomics, Ebersberg, Germany) by polymerase chain reaction (PCR) using GoTaq® Polymerase Master Mix (Promega, Mannheim, Germany) according to the manufacturer´s protocol.

Cloning was performed in two steps. First, the vector AAV-CAGGS-EGFP and the CMG-NTSR1 gene sequence were digested using appropriate restriction enzymes. Digestion products were run on an agarose gel and purified with gel extraction or Monarch PCR purification kits (NEB, Frankfurt am Main, Germany). Vector and CMG-NTSR1 were ligated with T4 DNA Ligase (NEB, Frankfurt am Main, Germany). 5 µl ligated vector DNA was transformed into 50 µl of competent Escherichia coli Top10 cells by incubation on ice for 2 min followed by heat shock incubation at 42°C for 30 s. After adding 200 µl of S.O.C. media (ThermoFischer Scientific (Invitrogen), Waltham, MA, USA) the cells were shaken at 37 °C for 30 min, and finally plated on LB agar containing 100 µg/ml ampicillin (Merck (Sigma), Darmstadt, Germany). Second, the vector containing the CMG-NTSR1 variant and the PCR-amplified mRuby3 were digested, ligated and transformed as described for the first cloning step, resulting in a vector containing the CMG-NTSR1-mRuby3-gene.

**2.2 Transfection of HEK293T FlpIn cells**

HEK293T FlpIn cells (ThermoFischer Scientific, Waltham, MA, USA) were cultured in high-glucose Dulbecco's Modified Eagle Medium (DMEM, ThermoFischer Scientific, Waltham, MA, USA) supplemented with 10% fetal bovine serum (FBS) (Biochrom, Berlin, Germany) and 1% penicillin/streptomycin (Biochrom, Berlin, Germany). 12-16 h prior to transfection, 1.5·10$^6$ HEK293T FlpIn cells were seeded into a six-well plate in 3 ml high-glucose DMEM supplemented with 10% FBS to be 70-90% confluent prior to transfection. Transfection of cells was accomplished by transcription activator-like effector nucleases (TALEN). A pair of TALEN carrying the Fok1 endonuclease target the left (L) and right (R) site of the human AAVS1 locus (the safe harbor site), respectively, which are 14-20 bp apart. The dimer formed by the two TALEN generates a double-strand break for introduction of the donor CMV-NTSR1-mRuby3-gene via homology-directed repair. The two TALEN plasmids hAAVS1_1L (Addgene # 35431) and hAAVS1_1R (Addgene # 35432) were a gift from Feng Zhang[20]. For transfection the donor plasmid and the TALEN plasmids were diluted in improved minimal essential medium (Opti-MEM, ThermoFischer Scientific, Waltham, MA, USA) according to the manufacturer's protocol, before 12 µl of Lipofectamine 2000 transfection reagent (ThermoFischer Scientific, Waltham, MA, USA) per well was added. This transfection mix was incubated for 5 min at room temperature before 300 µl was added dropwise to each well containing HEK FlpIn cells. Cells were selected with 1 µg/ml puromycin, and incubated at 37 °C under a 5% CO2 atmosphere over night before further use. Successful transfection of the cell line was verified by sequencing the PCR product of the genomic DNA fragment that contained the inserted CMV-NTSR1-

**2.3 Cell culture**

HEK293T FlpIn cells were maintained in complete cell culture media, i.e. Dulbecco's modified Eagle's medium (DMEM, Gibco) supplemented with 10% FBS (Biochrom) and 1% antibiotics (Penicillin/Streptomycin, Biochrom), and cultured at 37 °C. After 3 days cells were split and the cell culture medium was refreshed. Stable transfected HEK293T FlpIn were grown in complete cell culture media supplemented with 100 µg/ml hygromycin B (Roth, Karlsruhe, Germany).

**2.4 Microscopy**

Widefield imaging and structured illumination microscopy (3D-SIM) was performed on a Nikon N-SIM / N-STORM microscope[21-24] using 488 nm or 561 nm laser excitation. Appropriate SIM gratings were mounted for the Nikon 60x water immersion objective with N.A. 1.27 (CFI ) or the Nikon 100x silicon oil immersion objective with N.A. 1.35 (CFI SR HP lambda S 100XC Sil; temporary loan from Nikon, Germany), respectively. An additional 1.5-fold magnification lens was combined with the 2.5-fold magnification lens for the 3D-SIM mode when measuring with the 60x water immersion objective. Microscopic images were recorded by a cooled EMCCD camera (iXon DU-897 with EM gain set to "300", conversion gain 1x or 2.5x; Andor) using the Nikon "SIM488" (emission 500 nm to 540 nm) or "SIM561" (emission 570 nm to 640 nm) optical filter sets. SIM images of living cells were recorded at 23°C or 37°C, images of fixed cells were recorded at 23°C. Nikon analysis software was used for SIM image reconstruction or deconvolution, and for 3D visualization using the maximum intensity projection option.

For time-resolved anisotropy and FLIM imaging, a custom-designed confocal microscope with 3D piezo scanning system (Physik Instrumente) mounted on an Olympus IX71 and equipped with an 60x water immersion objective (UPLSAPO 60XW with N.A. 1.2, Olympus, Tokyo, Japan) or 60x silicon oil immersion objective (UPLSAPO 60XS2 with N.A. 1.3, Olympus) was used[25-28]. Ps-pulsed excitation was provided with 488 nm at 40 MHz (PicoTA 490, Picoquant) or with 561 nm at 80.6 MHz (SC-450, Fianium, UK), respectively. Up to 2 W of the NIR laser light from the Fianium supercontinuum laser were dumped to beam blocks (Thorlabs, USA) using 2 consecutive short pass filters with 900 nm cut-off wavelength. Subsequent wavelength selection of the pulsed laser was achieved using an tunable filter system with 14 nm bandwidth (TuneBox with filter VersaChrome HC 617/14, AHF, Tübingen, Germany) plus an additional narrow clean-up filter centered at 561 nm with 4 nm FWHM (561/4 BrightLine HC, AHF). Passing the 561 nm laser beam through a polarizing beam splitter cube (Thorlabs) resulted in the linear polarization required for FLIM and time-resolved anisotropy measurements.

Fluorescence photons were detected with up to four single-photon counting avalanche photodiodes (i.e. two SPCM-AQRH-14-TR APDs optimized for time-resolved measurements and two regular SPCM-AQRH-14 APDs, Excelitas, Canada) for simultaneous spectrally-resolved FLIM and time-resolved anisotropy measurements[27]. Four synchronized TCSPC cards (SPC-154, Becker&Hickl, Berlin, Germany) or one TCSPC card in combination with an 8-channel router (HRT-82, Becker&Hickl) recorded the photons. Adjusting the two APDs for FLIM and time-resolved anisotropy measurements required a careful shifting of the photon signals of one APD with ps time resolution that was achieved by an electronic ps delay generator ($PSD-065-A-MOD, Micro Photon Devices, Bolzano, Italy). A second electronic delay generator war used to amplify the weak trigger signals from the Fianium laser to synchronize the TCSPC electronics. FLIM and time-resolved anisotropy images were analyzed using the SPCImage software (Becker&Hickl), whereas intensity-based anisotropy images were analyzed using counter cards (National Instruments, USA) and custom Matlab scripts (Mathworks, USA)[29-38].

For live cell imaging, HEK293T cells were grown in microfluidic sample chambers with 170 µm cover glass bottom (µ-Slide VI[0.5], IBIDI, Planegg, Germany). These 6-channel slides could be connected to syringes using sterile Luer connectors, silicone tubing and injection ports (IBIDI). The height of a channel was 0.54 mm, and the total volume inside one channel was 40 µl according to the manufacturer. Buffer exchange and addition or transient exposure of neurotensin to NTSR1 in HEK293T cells was achieved with minimal volumes in these microfluidic sample chambers. We mechanically modified the sample holder of the heating incubator (Tokai Hit, Japan) used on the Nikon microscope to enable stable mounting of the larger IBIDI chambers.

# 3 RESULTS

**3.1 Imaging NTSR1-mNeonGreen and NTSR1-mRuby3 in a mixed population of HEK293T cells**

To evaluate expression, maturation of the fluorescent protein tags and correct localization of the NTSR1 mutants in the plasma membrane, living HEK293T cells were initially inspected by eye using a standard epifluorescence microscope. Afterwards we applied EMCCD-based image recording using a Nikon N-SIM / N-STORM microscope. Stable transfected HEK293T cells were grown in Dulbecco's modified Eagle's medium with 10% fetal bovine serum and appropriate antibiotics for three days in a microfluidic sample chamber (μ-slide VI$^{0.5}$, IBIDI) at 37°C. These chambers with 170 µm cover glass bottom are suitable for high-resolution microscopy and allow for building simple microfluidics based on Luer connectors, thin silicone tubing with an inner diameter of 1.6 mm, and sealed injection ports. The total volume inside the microfluidic channel was measured to be about 40 µl.

Both receptor mutants NTSR1-mNeonGreen and NTSR1-mRuby3 were mainly located at the plasma membrane after the HEK293T cells had been kept on PBS buffer supplemented with $Ca^{2+}$ and $Mg^{2+}$, called PBS+ in the following (DPBS, ThermoFischer Scientific), for 1 h at 37°C. The PBS+ buffer replaced the growth medium and was used both to maintain cell adhesion and to reduce fluorescence background. Figure 1 shows a mixed population of HEK293T cells either expressing NTSR1-mNeonGreen or NTSR1-mRuby3. Images were recorded first with 561 nm excitation for NTSR1-mRuby3 followed by 488 nm excitation for NTSR1-mNeonGreen. The apparent brightness of mRuby3 was found to be significantly lower than for mNeonGreen. Also the photostability of mRuby3 was lower and resulted in faster photobleaching at moderate laser power. Accordingly, 3D-SIM imaging that is based on recording 15 frames for one image was possible for a few bright cells. The SIM image classification process often resulted in a poor data score due to failed identification of the SIM grid position in the image. Instead of a full SIM reconstruction these images were deconvolved only.

We found homogenous intensity distributions in the membrane indicating both a reasonably high number of NTSR1 (in contrast to spatially separated individual molecules) and a diffusive mobility of the receptors during the recording of the image, i.e. motion blur. Inside the HEK293T cells and near the plasma membrane we also noticed some bright spots that were attributed to either the endoplasmatic reticulum (ER) or to some endosomes. Spectral cross talks, i.e. mNeonGreen fluorescence excited at 561 and detected between 575 nm and 625 nm as well as mRuby3 fluorescence excited at 488 nm and detected between 500 nm and 550 nm, were found to be small (see Fig. 1).

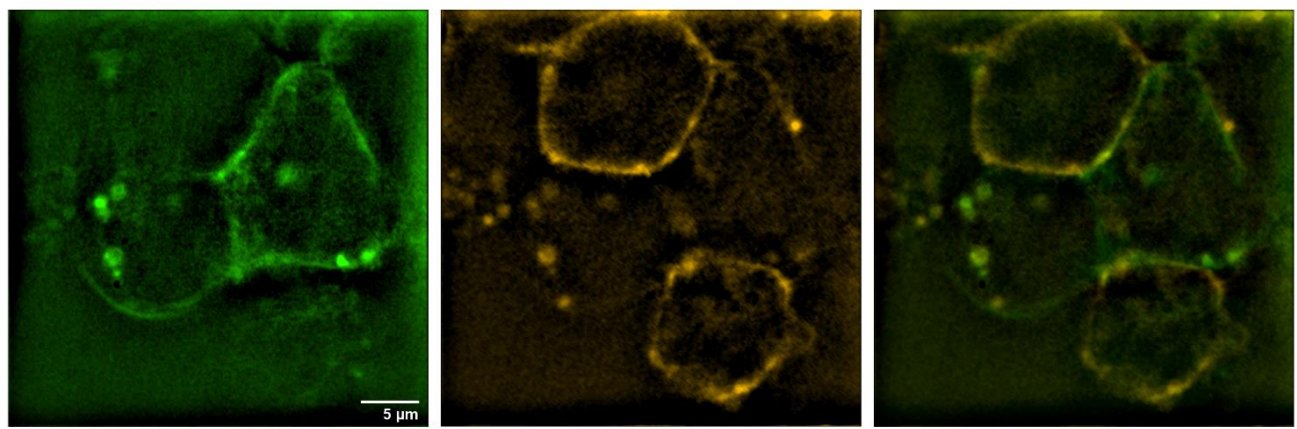

**Figure 1**: Deconvolved false-colored images of a mixed population of living HEK293T cells expressing either NTSR1-mNeonGreen or NTSR1-mRuby3. **Left**, NTSR1-mNeonGreen is expressed in the membranes of two cells in the center of the image (excitation: 488 nm). **Middle**, NTSR1-mRuby3 is expressed in the membranes of two other cells at the upper and lower part of the image (excitation: 561 nm). **Right**, overlay of the two images, with the NTSR1-mNeonGreen image intensity contribution set to 40%. Images were recorded as 15 frames for 3D-SIM at 23°C using the Nikon 60x water immersion objective with combined 1.5x and 2.5x magnification.

## 3.2 FlAsH labeling of NTSR1-mRuby3 in HEK293T cells

The NTSR1-mRuby3 mutant contained also a tetracysteine motif by replacing the original residues $_{290}$IEPGRV$_{295}$ with $_{290}$CCPGCC$_{295}$ in the ICL3 of the receptor. We evaluated the accessibility of this tetracysteine motif by labeling NTSR1-mRuby3 with the bisarsenical fluorescein dye FlAsH in stably transfected HEK293T cells. For labeling at 37°C, we added 500 nM FlAsH (ThermoFisher Scientific) for 1 h in the presence of 12.5 µM EDT in PBS+ to the HEK293T cells, and removed the unbound dye afterwards by incubating the cells for 10 min with 250 µM EDT followed by washing with PBS+ buffer.

Figure 2 shows HEK293T cells expressing NTSR1-mRuby3 and labeled with FlAsH. We started SIM imaging of mRuby3 first using 561 nm excitation. Subsequently we excited FlAsH with 488 nm for SIM imaging. Fluorescence intensities were low for both FlAsH and for mRuby3, and the S/N ratio was only good enough for deconvolution of the images but not for SIM image reconstruction. The plasma membrane of the cells appeared in both deconvolved fluorescence images, but the brighter spots inside the cells were apparently not colocalized in the overlaid image in Fig. 2. To reveal the specific binding of FlAsH to the engineered tetracysteine motif on NTSR1-mRuby3 we will have to apply FRET microscopy instead of sequential colocalization. FRET imaging avoids motion blur due to the permanent shape changes of living cells. Nevertheless, the small spectral cross talk of mRuby3 into the FlAsH detection scheme allows to interpret the fluorescence intensities on the membrane using 488 nm excitation as related to FlAsH staining.

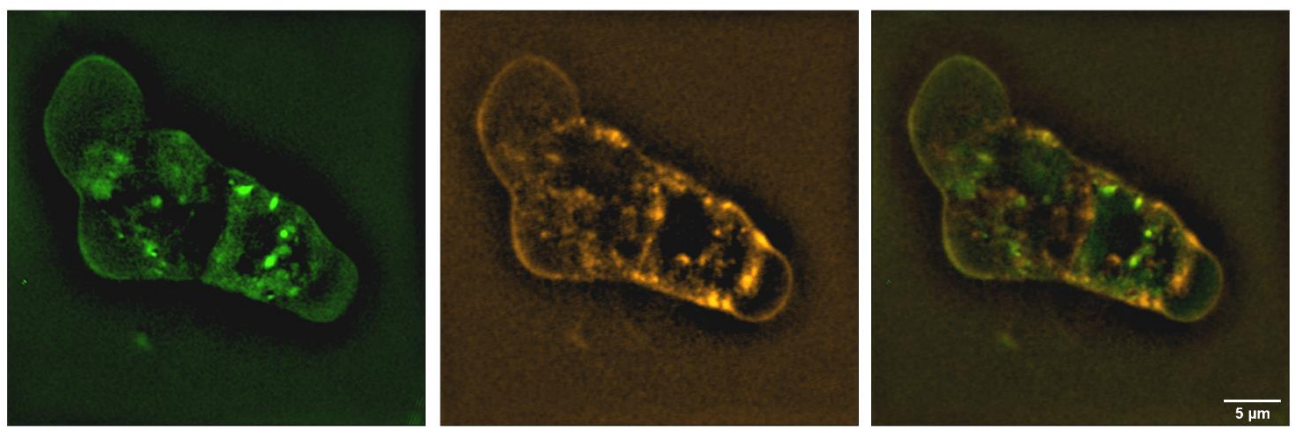

**Figure 2:** Labeling NTSR1-mRuby3 with FlAsH in HEK293T cells. **Left**, FlAsH excited with 488 nm. **Middle**, NTSR1-mRuby excited with 561 nm. **Right**, overlay of the two images, with the FlAsH image intensity contribution set to 40%. Images were recorded as 15 frames for 3D-SIM at 23°C using the Nikon 60x water immersion objective with 2.5x magnification only. Deconvolved images are shown with enhanced image contrast.

## 3.3 Time-lapse imaging of NTSR1-mRuby3 in living HEK293T cells at 37°C

Next we probed the activation of the NTSR1 mutants by addition of its ligand neurotensin. Neurotensin is a short peptide composed of 13 amino acids that binds with its C terminus into the binding site on NTSR1. However, only residues 8 to 13 are required for activating the receptor. We have synthesized the essential C-terminal part of neurotensin and added a cysteine for possible labeling with fluorophores based on maleimide chemistry[8]. The unlabeled version of the peptide CKPRRPYIL was used here.

To transiently expose NTSR1-mRuby to its ligand we constructed a microfluidic flow chamber with sterile IBIDI components within the heated incubation system of the Nikon microscope. One side of the flow chamber (containing adhered living HEK293T cells at 37°C) was connected to a 2 ml syringe using 5 cm of a thin silicon tubing with 1.6 mm inner diameter. This part was filled with 1 ml PBS+ buffer without air bubbles. The other outlet of the microfluidic chamber was connected *via* a 5 cm silicon tubing to an sterile injection port with septum. Following the injection port, a silicon tubing of 10 cm length was attached, and this part was also filled with PBS+ without any air bubbles. To apply neurotensin to the cells afterwards, the injection port could be loaded with 200 µl of the ligand solution which pushed the

PBS+ buffer in the silicon tubing away from the cells, i.e. using the 10 cm silicon tubing as a reservoir. When pulling the syringe, the neurotensin was moved across the cells, and when pushing the syringe back PBS+ buffer was used to wash away the concentrated neurotensin solution. We achieved a continuous microscopy recording of the same HEK293T cells during exposure with neurotensin and followed by washing with PBS+ avoiding strong mechanical distortions of the sample chamber.

Because of the limited photostability of mRuby3 with a reported quantum yield of only 45%[17], we measured image intensity losses for different laser powers by time-lapse imaging. Shown in Fig. 3 are individual frames of a 10 min recording using the widefield mode of the Nikon N-SIM microscope with 561 nm excitation. The EMCCD exposure time per frame was set to 3 s, with a duty cycle of 5 s per frame.

The false-colored intensities of NTSR1-mRuby3 in the membrane remained almost constant indicating only small losses. Membranes appeared to be continuously stained with similar pixel intensities corresponding to 'light blue' values in the Nikon lookup table "*green fire blue*". Inside the cells, for example the one on the upper left part of the image, we found few bright spots as described above. We noticed that the cells were slowly moving and changed their shape within 10 min, but stayed in the field of view.

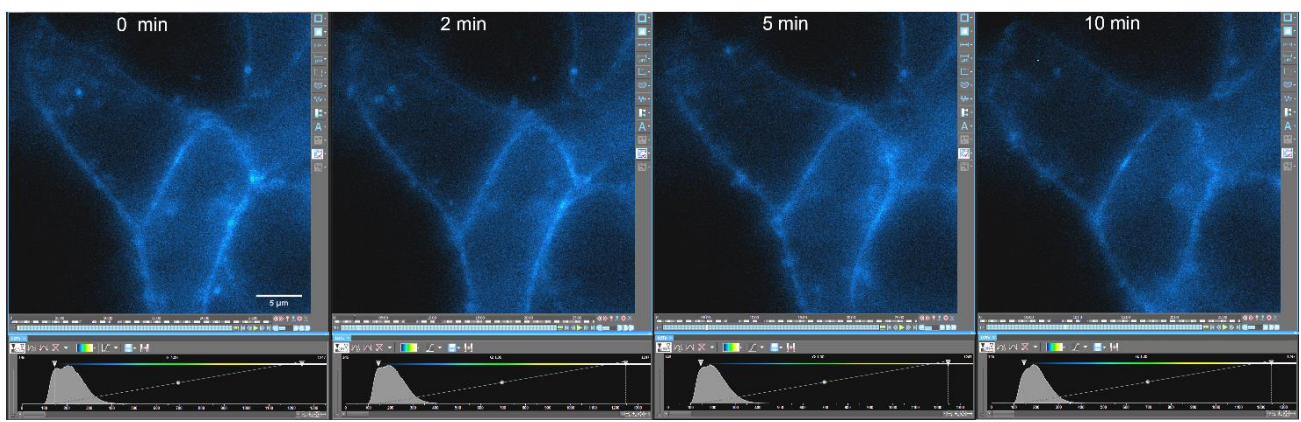

**Figure 3**: Time-lapse widefield imaging of NTSR1-mRuby3 in HEK293T cells at 37°C. The Nikon 100xSi objective was used in combination with 2.5x magnification. Excitation: 561 nm, 3 s exposure time per frame. Intensities are false-colored with the same settings using the Nikon lookup table "green fire blue" shown in the lower panels. Screen shot images were modified for enhanced image contrast. Scale bar is 5 µm (left image).

### 3.4 Time-lapse imaging of activated NTSR1-mRuby3 in living HEK293T cells following neurotensin exposure at 37°C

We continued the recording of the HEK293T cells shown above, and added 50 µM of the shortened neurotensin derivate, i.e. the peptide CKPRRPYIL, for transient exposure. Neurotensin exposure was estimated to last for several seconds while the chamber with HEK293T cells was flushed with the neurotensin solution. Flow was achieved by manually pulling the syringe attached to the IBIDI microfluidic chamber.

Figure 4 shows the time series over 15 min. Already 1 min after exposure to neurotensin the homogeneous distribution of NTSR1-mRuby3 in the membranes started to change into a more heterogeneous appearance. After 2 min we identified slightly larger spots of NTSR1-mRuby3 in the membrane, but with similar intensities ("light blue") as before addition of neurotensin. After 3 min the clustering of NTSR1-mRuby3 resulted in the first significantly brighter spots as indicated by a "green" pixel color. However, some parts of the membranes still showed an almost homogeneous fluorescence intensity corresponding to a homogeneous NTSR1-mRuby3 distribution. The HEK293T cells changed their shape during the time series either due to drifting of the focal plane or due to the shear force associated with the flow across the cells.

About 4 min after neurotensin exposure the bright spots of clustered NTSR1-mRuby3 (i.e. pixels colored "green") increased in size. These bright spots of aggregated receptors remained localized and diffused only slowly. The parts of the membranes with a homogeneous distribution of NTSR1-mRuby3 diminished in the following minutes. After 10 min, only small areas of the membranes still contained a homogeneous distribution of NTSR1-mRuby3 but almost all of the receptors were clustered. As a control for the amount of photobleaching of mRuby3 we compared the fluorescence intensities of the cells that were illuminated continuously over 15 min after exposure to neurotensin. An obvious loss of fluorescence was observed in Fig. 4 (see image at the lower right). However we could not exclude that the mechanical drift of the focal plane contributed to the loss of fluorescence as well.

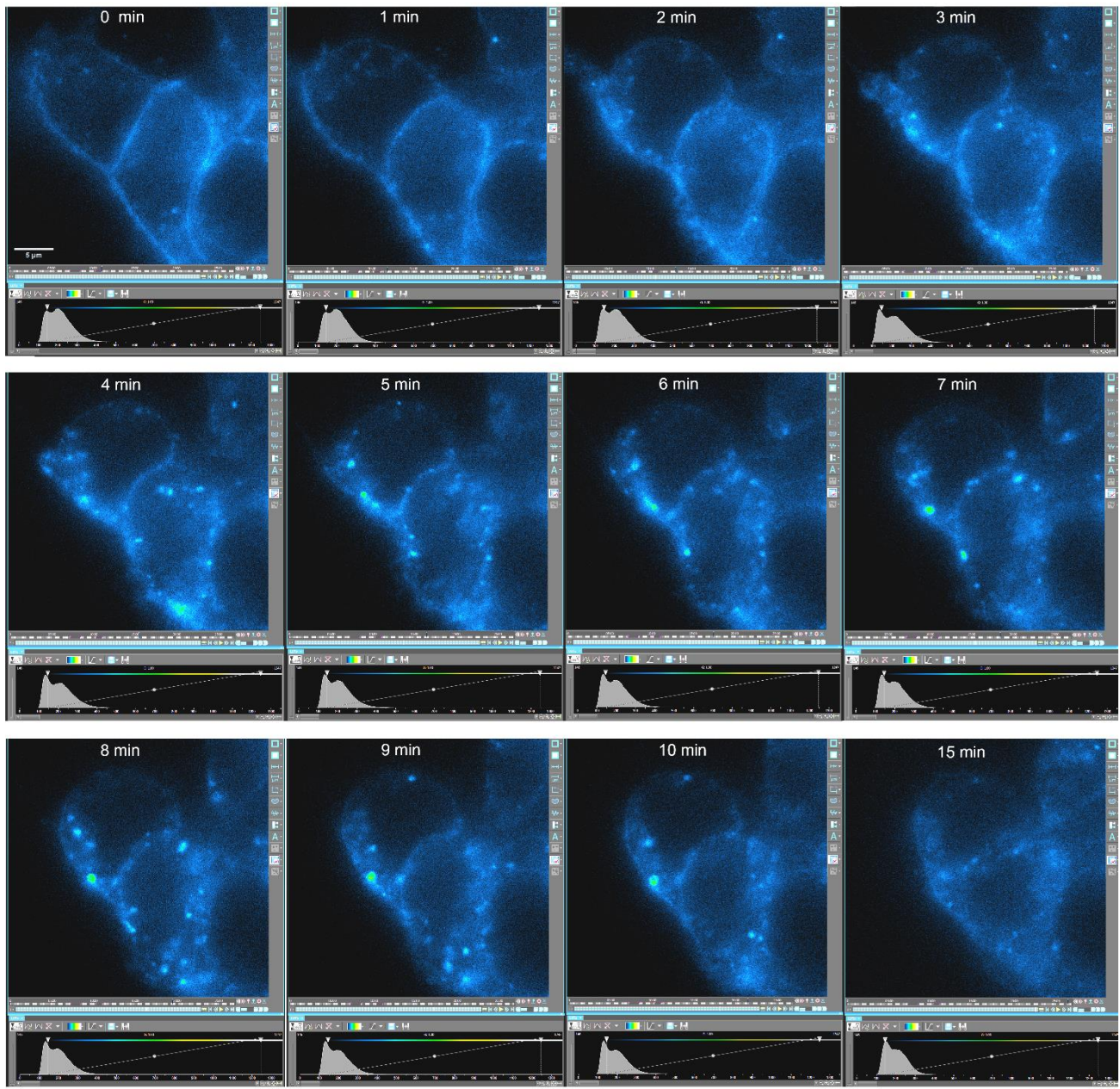

**Figure 4**: Time-lapse widefield imaging of NTSR1-mRuby3 in HEK293T cells after transient exposure to 50 µM neurotensin at 37°C. The Nikon 100xSi objective was used in combination with 2.5x magnification. Excitation 561 nm, 3 s exposure time per frame. Intensities are false-colored with the same settings using the Nikon lookup table "green fire blue" as shown in the associated lower panels. Screen shot images were modified for enhanced image contrast. Scale bar (upper left image) is 5 µm.

The changes of cellular shapes and membrane movements between neighboring HEK293T cells over the time course of 15 min limited any kinetic analysis of the formation of individual NTSR1-mRuby3 clusters as well as the possibility to estimate the number of receptors within such clusters.

### 3.5 FLIM and anisotropy imaging of NTSR1-mNeonGreen in HEK293T cells at 21°C

The neurotensin-induced clustering of NTSR1 was also observed similarly for the NTSR1-mNeonGreen mutant in HEK293T cells at 37°C (data not shown here). However, fluorescence intensity imaging did not provide a measure of the oligomerization status of NTSR1 before and after exposure to its ligand. Therefore we applied confocal FLIM and anisotropy microscopy of NTSR1-mNeonGreen expressed in HEK293T cells using the IBIDI microfluidic chamber as described above for addition or exposure of neurotensin to the same set of cells, respectively. Time-resolved confocal microscopy was achieved using a slow 3D piezo scanner with 4 ms pixel dwells but without a temperature-controlled incubation system, i.e. the images were recorded at 21°C using pulsed excitation at 488 nm.

Figure 5 shows FLIM images and the calculated static anisotropies per pixel of NTSR1-mNeonGreen in HEK293T cells before and about 16 min after exposure to neurotensin.

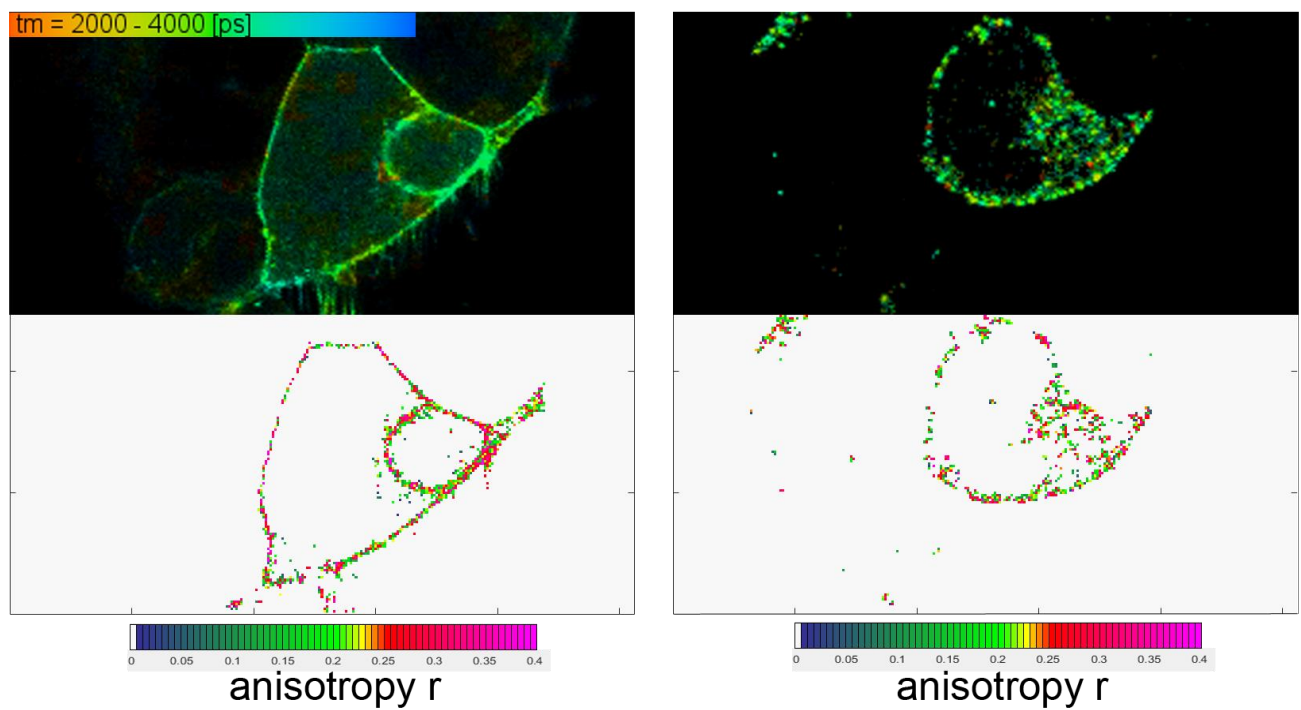

**Figure 5**: Confocal FLIM and static anisotropy images of HEK293T cells expressing NTSR1-mNeonGreen before (left) and 15 min after exposure to 500 nM neurotensin (right) at 21°C. Pulsed excitation with 488 nm at 40 MHz, emission detected between 500 nm to 570 nm. The fluorescence lifetimes in the two FLIM images (upper panels) are color-coded with the same mean lifetime $\tau_m$ scale in ps shown on the upper left. The corresponding static anisotropy images (lower panels) are color-coded using the same anisotropy scale bar as shown below the panels. The size of the four images was 256 x 128 pixel corresponding to 100µm x 50 µm, using an 60xW objective without additional magnification.

Simultaneous to the FLIM image the fluorescence anisotropy image was measured using the intensity per pixel information. Polarized emission was recorded by the two APDs and an apparent steady-state anisotropy *r* was calculated for each pixel using

$$r = \frac{I(parallel) - I(perpendicular)}{I(parallel) + 2*I(perpendicular)} \tag{1}$$

without further corrections for a measured G factor for the slightly different detection efficiencies of the two channels, and, importantly, without corrections for the depolarization caused by the high numerical aperture of 1.2 of the 60x water immersion objective.

The FLIM images showed membrane staining with a fluorescence lifetime of 3 ns (colored bright green), i.e. the expected lifetime of mNeonGreen fused to NTSR1 ($\tau_m$ = 3.15 ns for purified NTSR1-mNeonGreen[8]). 15 min after exposure to neurotensin at 21°C the receptor had clustered, but the fluorescence lifetime of NTSR1-mNeonGreen did not change. A significant shift of the focal plane was observed when comparing both FLIM images in Fig. 5, i.e. the mechanical stability of the piezo stage on our microscope was not perfect.

In contrast, the distribution of anisotropy values for NTSR1-mNeonGreen in the membrane before and after exposure to neurotensin changed clearly.

Before addition of neurotensin (lower left image in Fig. 5), many pixels appeared in pink to red colors corresponding to high fluorescence anisotropy values of $r > 0.25$. These high anisotropies likely indicate monomeric NTSR1-mNeonGreen without direct interaction with other NTSR1-mNeonGreen molecules. However, we also found some yellow to green pixels in the membrane, i.e. pixels with anisotropy values of $r < 0.25$. These lower anisotropies likely correspond to dimeric (yellow) or higher oligomeric NTSR1-mNeonGreen receptors that exhibit homoFRET between their mNeonGreen tags.

After exposure to neurotensin (lower right image in Fig. 5), more pixels appeared as yellow to green pixels in the membrane, i.e. pixels with low anisotropy values of $r < 0.25$. Some pixels were found in dark green, i.e. with anisotropy values of $r < 0.1$. A qualitative comparison of similar areas of the cell membrane indicated that activation by neurotensin caused (1) formation of receptor clusters represented by individual spots instead of lines of pixels, (2) an increase in fluorescence intensities for clusters of aggregated NTSR1-mNeonGreen (intensity data not shown here), (3) no changes of the fluorescence lifetime, but (4) a shift from higher to lower anisotropies due to homoFRET between identical fluorophores at shorter distances below 10 nm.

## 4 DISCUSSION

We have generated a plasmid for the human G protein-coupled receptor NTSR1 with a C-terminal fusion of either mRuby3 or mNeonGreen and a tetracysteine motif in the intracellular loop 3 (ICL3) of the receptor. The plasmid was used to stably transfect HEK293T FlpIn cells[39] for live cell imaging. NTSR1 was localized predominantly at the plasma membrane of the HEK293T cells after replacing the growth medium by PBS buffer supplemented with $Ca^{2+}$ and $Mg^{2+}$ and incubation for 1 h at 37°C.

SIM imaging of living HEK293T cells revealed that NTSR1-mRuby3 was produced at low concentrations resulting in low fluorescence intensities. The receptor was distributed homogeneously in the membrane before activation by a shortened derivative of neurotensin. Inside the cytosol and near the membrane we found few bright fluorescent spots that were attributed to the ER or to endosomes. Preliminary labeling experiments with FlAsH for the tetracysteine at ICL3 of NTSR1-mRuby3 resulted in membrane staining but also in labeling of sub-compartments of the HEK293T cells. Optimization of the FlAsH labeling protocol in the presence of an appropriate dithiol component is required before intramolecular FRET measurements between FlAsH as the FRET donor and mRuby3 as the FRET acceptor are possible. Despite the known photophysical limitations of FlAsH (lower quantum yield at physiological pH, limited photostability and blinking), this fluorophore has been used also for ultrasensitive single molecule detection[16]. Similarly, mRuby3 with a quantum yield of 45% even at low excitation power[17] and with other photophysical dark states (see A. Dathe et al. Proc. SPIE 10884, *in press* [2019]) could be replaced by a brighter fluorescent protein, for example mScarlet[40], to enable FRET-based analysis of conformational changes of NTSR1 upon activation by its ligand neurotensin.

Functionality of the mutant NTSR1-mRuby3 was demonstrated in living HEK293T cells by induced redistribution of the receptors following exposure to neurotensin. Within a few minutes after addition of neurotensin, NTSR1 started to cluster in the membrane. While the initial brightness of these NTSR1-oligomers was similar to the brightness before addition of neurotensin, the fluorescence intensities of the clusters and their apparent sizes increased after a few minutes. Extended observation times resulted in a loss of intensity due to photobleaching of mRuby3.

Confocal FLIM imaging confirmed that NTSR1-mRuby3 was diffusing within the in membrane according to its fluorescence lifetime of $\tau \sim 3$ ns. The oligomeric status of the receptor could not be assessed from the lifetime, but was revealed by the simultaneous recording of the fluorescence anisotropy in each pixel. Steady-state anisotropy images revealed that NTSR1-mRuby3 was mostly monomeric according to anisotropy values of $r > 0.25$. However, we also detected pixels of the membrane area with lower anisotropies indicating a dimeric NTSR1 because of homoFRET between the mRuby3 fluorophores. The oligomerization status of the receptor changed towards larger aggregates with $r < 0.2$ after exposure to neurotensin, and we found more pixels of the membrane area with low anisotropies after 10 to 15 minutes. Therefore we expect that NTSR1 might exist in a dynamic equilibrium between monomer and dimer in the absence of its ligand as observed *in vitro*[7], but NTSR1 aggregates into larger clusters after activation by neurotensin.

Future imaging of NTSR1 activation and its conformational dynamics in living cells will have to combine both imaging modalities shown here, i.e. video-rate high resolution imaging of several cells in the field of view with functional imaging based on fluorescence lifetime, anisotropy (homoFRET) and FRET between different fluorophores. One very recent approach comprised the combination of confocal superresolution microscopy, i.e. image scanning microscopy (ISM), and FLIM[41]. Depending on the brightness of the fluorophores on NTSR1, i.e. the number of photons per pixel and time bin, ISM-FLIM can be extended to add a second detection channel for homoFRET or for time-resolved anisotropy recording, respectively. Alternatively, STED microscopy in combination with FLIM and anisotropy measurements might be applied to unravel the spatial and functional dynamics of this human G protein-coupled receptor NTSR1 in living cells at work.

**Acknowledgements**

The authors thank all members of our research groups who participated in various aspects of this work, from genetics and biochemistry support to cell culture. We are grateful for the loan of the 100xSi superresolution objective by Nikon, Germany. Financial support for the Nikon N-SIM / N-STORM superresolution microscope by the State of Thuringia (grant FKZ 12026-515 to M.B.), for A.W. through LOM funds 2017 by the Jena University Hospital (to M.B.) and in part by the Deutsche Forschungsgemeinschaft DFG in the Collaborative Research Center/Transregio 166 "ReceptorLight" (project A1 to M.B.) is gratefully acknowledged.